# Similarity between the primary and secondary air-assisted liquid jet breakup mechanism

Y.J. Wang, Kyoung-Su Im, K. Fezzaa

*X-Ray Science Division, Argonne National Laboratory,* Argonne, IL 60439

e-mail: yujie@aps.anl.gov

In this letter we report an ultrafast synchrotron x-ray phase-contrast imaging study of the primary breakup mechanism of a coaxial air-assisted water jet. We demonstrate that there exists great analogy between the primary breakup of the water jet with its secondary breakup on a quantitative level. Especially, a transition from a ligament-mediated breakup to a membrane-mediated breakup is identified around effective Weber number ~ 13 for the surface undulations induced primary breakup. This observation reveals the intrinsic connections of these two breakup regimes and has significant implications on the unified theoretical approach in treating the breakup process of high speed liquid jet. Moreover, since the results are obtained in a highly turbulent breakup regime, it suggests that aerodynamic Weber number plays a very important role in determining the breakup process.

Spray generation by a low-speed liquid flow exposed to a surrounding co-flowing high-speed gas stream is of great importance for industrial applications, such as fuel atomization in rocket engine, spray painting (1), and inkjet research. However, due to its complexity and transient nature, current knowledge of the breakup process is very limited and a consensus on the breakup mechanism is still long way to be reached. It is generally believed the breakup process consists of removing liquid mass from the liquid jet surface to form droplets and breaking up these droplets subsequently. These two distinctive sub-processes are conventionally termed as the primary and secondary breakup and treated independently. Marmottant *et al.* (2, 3) have studied the primary breakup of a coaxial air-assisted water jet breakup by visible light imaging. A unique ligament-mediated primary breakup model is proposed. A subsequent self-similarity convolution of breakup process is used to estimate the eventual droplet size distribution (2, 3). In the study, ligament-mediated breakup seems to be the dominant primary breakup mechanism (2, 3). However, the air speed range studied in the work does not include high air speed cases. Additionally, the nozzle utilized in the experiment was specially designed to reduce the initial air and liquid turbulence, so how general this ligament-mediated breakup can be applied to higher air velocities remains questionable. Varga *et al.* (4) studied a small-diameter liquid jet exposed to a very high-speed air stream and observed that there exist some similarities between the liquid jet primary breakup and that of the droplet breakup qualitatively (10, 11). In this letter, we report a detailed study of the primary breakup process of a coaxial air-assisted water jet over a wider air speed range in turbulent breakup condition by ultrafast synchrotron x-ray phase-contrast imaging. By exploiting the unprecedented spatial and temporal resolutions offered by this new technique, we have established the close analogy between the primary breakup of the liquid jet with its secondary breakup on a quantitative fashion. The ligament-mediated breakup remains low air velocity approximations.

The coaxial air-assisted water spray system used in this experiment is a commercial paint spray gun. It has a central circular orifice with diameter of 1.1 mm. The coaxial air annular orifice has an inner diameter of 2.48 mm and outer diameter of 3.90 mm. The water emanates from the central orifice with considerably lower speed as compared to the surrounding co-flowing air stream. By fixing the water flow speed, the breakup phenomenon is studied in a wide air speed range. We obtain the cross-section averaged water and air nozzle exit speeds independently by measuring the water flow rate and by a commercial air velocity analyzer (SEAVA®). With water flow speed set at 8.3 m/s, the liquid Reynolds number ($Re = V_l D_l / v_l$) is 6000, thus highly turbulent, where $V_l$ denotes the water exit velocity, $D_l$ is the water jet diameter and $v_l$ is the water dynamic viscosity. The jet Weber number ($We = \rho_g (V_g - V_l)^2 D_l / \sigma$) is also defined, in which $\rho_g$ denotes the air density, $V_g$ the air exit velocity, and $\sigma$ is the water surface tension.

The breakup process is studied by ultrafast x-ray phase-contrast imaging at 32-ID undulator beamline of the Advanced Photon Source (APS) at Argonne National Laboratory. The "white beam" (unfiltered energy spectrum) x-ray radiation from the undulator delivers the necessary x-ray flux for the temporal and spatial resolutions of this experiment. The gap of the undulator is set to be 31 mm which corresponds to a 13.3 KeV first harmonic x-ray energy. At this gap setting, the image contrast is dominated by phase contrast instead of absorption contrast (5, 6). The x-ray beam has a source size of 15 μm (vertical, V) x 240 μm (horizontal, H) and a sample-to-source distance of 40 m. The transmitted and diffracted x-rays through the jet are converted to visible lights by a LYSO:Ce scintillator crystal placed at 750 mm downstream and imaged with optical lenses (5x, NA=0.14) onto a fast sensiCam CCD camera. The CCD camera has a 1024 x

1280 pixel-array chip with individual pixel sizes of 6.67 x 6.67 $\mu m^2$; the effective size of each pixel after optical magnification is 1.25 x 1.25 $\mu m^2$. The effective imaging resolution is around 2 $\mu m$ with the effective field of view around 1.37 x 1.71 $mm^2$.

The breakup process is studied by a combination of single-frame-single-exposure and single-frame-multiple-exposure imaging schemes. In the single-frame-single-exposure scheme, the snapshots of the breakup process were taken with a single 472 ns x-ray pulse. To catch the breakup dynamics, the single-frame-multiple-exposure scheme is utilized instead. This multiple-exposure scheme takes advantage of the repetitive temporal structure of the x-ray pulses generated from the synchrotron storage ring and as a result no high repetition-rate camera is needed. The exposure time for each single pulse is still 472 ns with inter-exposure of 17.4 $\mu s$. Exploiting this scheme, we can catch the breakup dynamics in a single image frame.

In Fig.1, the 472 ns single-frame-single-exposure x-ray phase contrast images of the water jet at different air speeds are plotted. The image frames took at different nozzle distances were spliced together to give an overall view of the breakup process. Since the effective field-of-view of the imaging system is just slightly bigger than the jet diameter, to capture the breakup process within a single frame, only half of the jet is imaged.

Qualitative differences of the breakup at different air speeds are easily recognizable. The high-speed air stream induces surface undulations. Since the liquid jet and the air stream are both turbulent, these surface undulations occur much more chaotically as compared to previous studies (2, 3). The surface undulations develop earlier and with bigger amplitudes as the air speeds increase and eventually cause the water mass disintegration from the main jet. It is generally observed that in the relative high air speed range, these surface undulations are responsible for the breakup process as compared to low air-speed cases when the water jet breaks up as a whole.

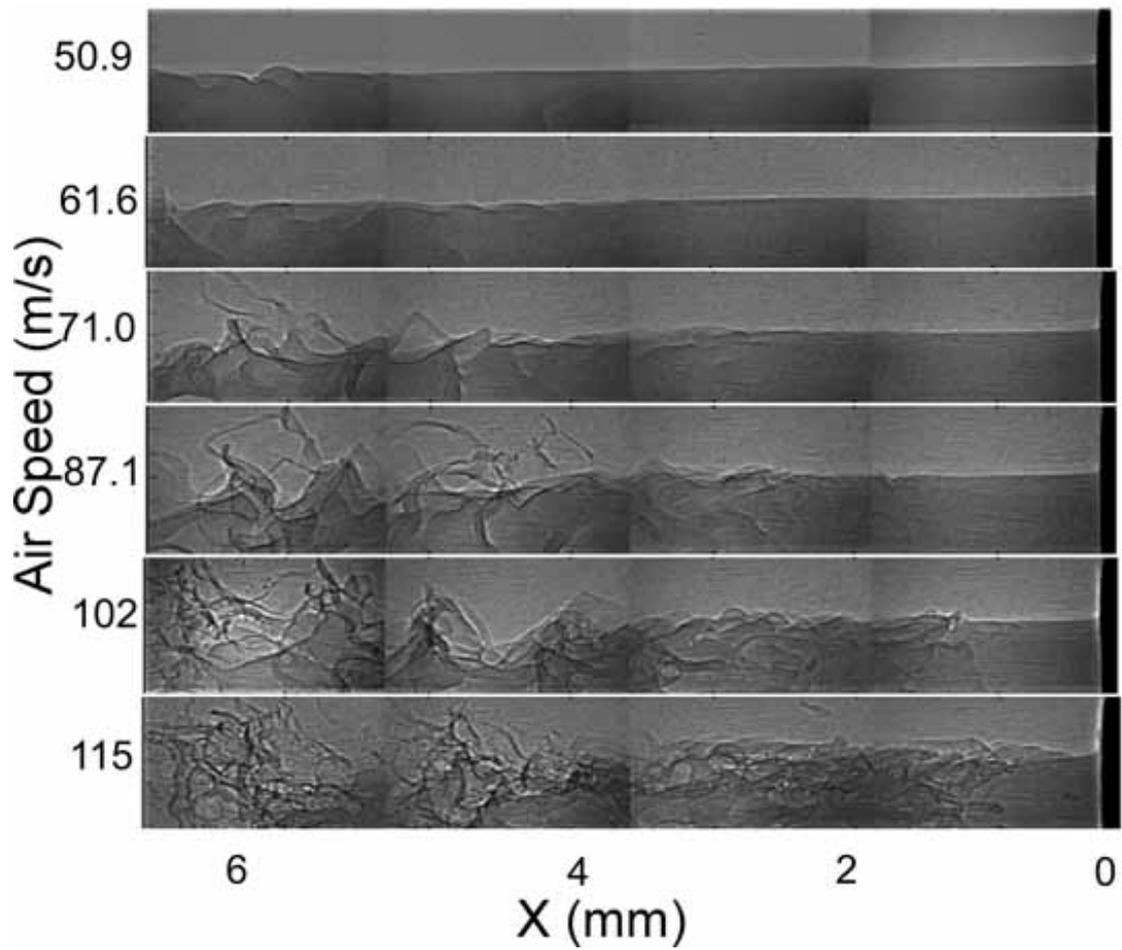

Fig. 1. Phase contrast images of the water jet with increasing air velocities. The jet is emanated from the right sides.

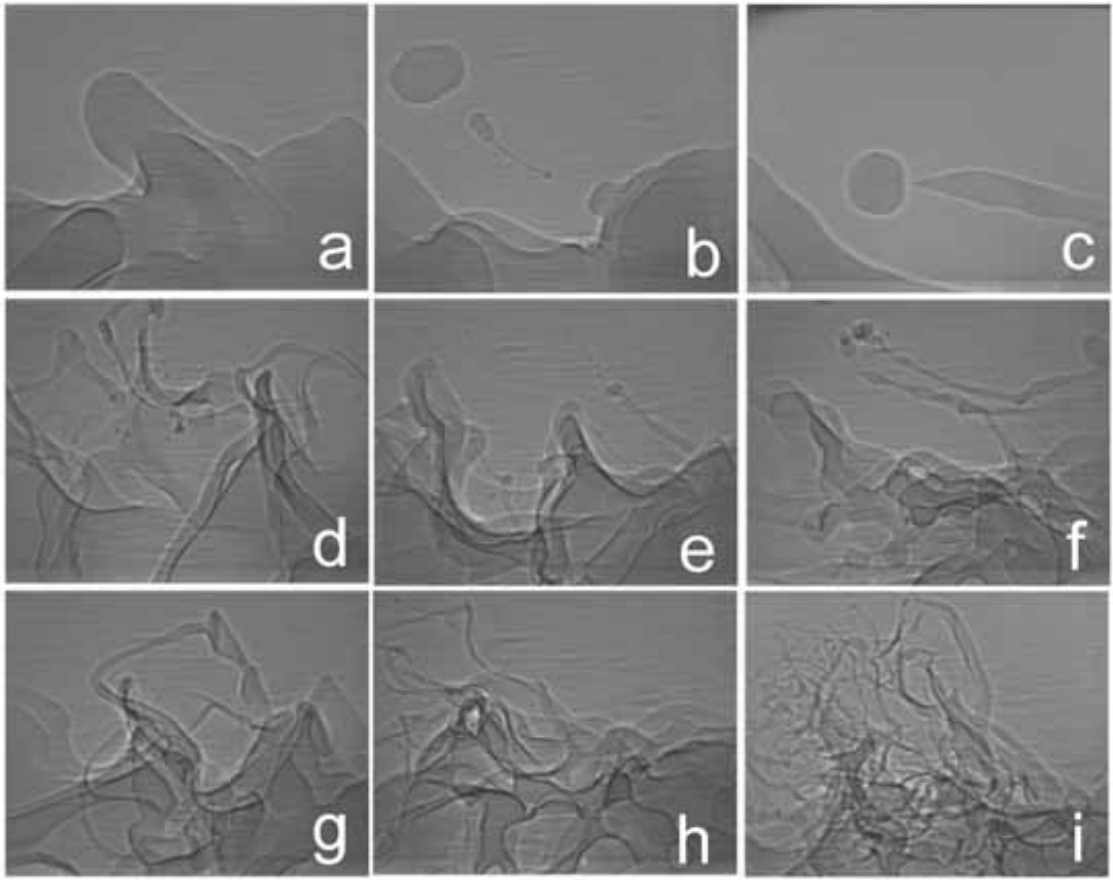

Fig. 2. Characteristic breakup images for increasing Weber numbers. (a), (b) $V_g$ = 50.9 m/s; (c) $V_g$ = 52.9 m/s; (d), (e), (f) $V_g$ = 71.0 m/s; (g) $V_g$ = 87.1 m/s, (h), (i) $V_g$ = 102 m/s. The red lines indicate the procedures adopted to estimate the breakup volume $V_{break}$ for both ligament-mediated and membrane-mediated breakups.

To understand the breakup process in close proximity, we carefully zoom in on the surface undulations by taking their snapshots and following their dynamics. The breakup process falls into two distinctive scenarios when $V_g$ < 60 m/s and $V_g$ > 60 m/s.

When $V_g < 60$ m/s, the surface undulations are amplified as they move downstream and form rod-like structure protruding into the air (Fig. 2a). Infrequently, the breakup happens in which a big lump of liquid mass disintegrates from the jet, as shown in Fig. 2b when $V_g = 50.9$ m/s. The retracting neck and capillary waves excited at both the tip and the base can be clearly seen. Satellite droplets generated by nonlinear effects are much smaller in sizes as compared with the main droplet.

When the air speeds are further increased, the surface rods are stretched more into the fast air stream to form long ligaments and break up subsequently. It is worthwhile to focus on the breakup dynamics of the ligaments. As is obvious from Fig. 3a-b when $V_g = 57.6$ m/s, the ligaments are stretched longer into the air environment by aerodynamic forces. The capillary instabilities in the middle of the ligament are suppressed and the ligament breaks off through an "end pinching" mode as suggested by Stone (7), the "end-pinching" happens at the neck between the end blob and the main ligament. The "end-pinching" time is determined by the capillary time $t_c = (\frac{\rho_l r^3}{\sigma})^{1/2}$ in low speed jet breakup (3), where $\rho_l$ is the liquid density, $r$ is the blob radius. It turns out that this criterion also applies to our high-speed breakup process. As shown in Fig. 3a, the end blob has a radius around 80 μm. The corresponding capillary time is around 55 μs. Indeed from the image, we can observe the breakup is almost complete after only three consecutive exposures. However, once the pinch-off of the blob from the ligament is completed, the aerodynamic forces acting on the ligament cease to exist and the ligament retracts as a result. The remaining ligament will break up owing to capillary instabilities. Interestingly, the dominant capillary instabilities remain the same as before the pinch off. In this breakup scenario, the breakup process generates more uniform size droplets.

The breakup scenario will drastically change when the air speed beyond 60 m/s. Instead of forming ligaments, the preformed surface undulations were blown into two-dimensional membrane structures with thick rims. The membrane structures continue to expand in the wind due to aerodynamic forces and eventually burst. As a result, the rim collects most of the liquid mass and breaks up in a secondary step due to capillary instabilities.

Fig. 2d-f illustrate the snapshots of different stages of the membrane-mediated breakup when $V_g$ is 61.6 m/s, and Fig. 3c-d illustrates in detail the membrane-mediated breakup process through the multiple-exposure imaging technique.

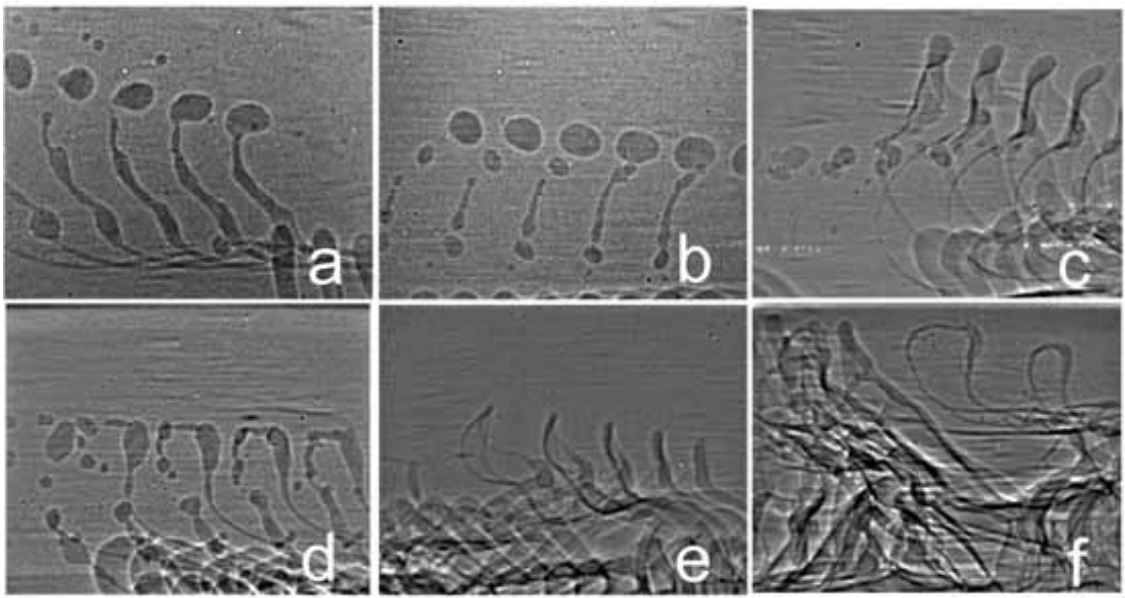

Fig. 3. Single-frame-multiple-shots of the breakup dynamics. (a), (b) $V_g$ = 57.6 m/s; (c), (d) $V_g$ = 61.6 m/s  (e) $V_g$ = 71.0 m/s ; (f) $V_g$ = 87.1 m/s. Images are earlier in time at the right side.

In Fig. 3b, it is observed that after the bursting of the membrane, the remaining rim is further stretched in the air stream. It breaks up in very similar fashion as the ligaments due to capillary instabilities.

Very high air speed cases are more difficult to study since the breakup happens in a more chaotic fashion. As shown in Fig. 2g-i, when the air speed is above 87 m/s, the breakup process is characterized as possessing large number of fine threads. However, the membranes can still be seen with much thinner rims with the dynamics captured in Fig. 3f. It is natural to presume that the generation of fine threads is still owing to the breakup of thinner membranes.

In total, the surface-undulation-induced-breakup process displays a rich zoology of phenomena as possessing both ligament- and membrane- mediated breakups. We have to emphasize that these types of breakups are different from those suggested by Chigier *et al.* (13) in which the jet as a whole breaks up through the Rayleigh and membrane breakup. In our case, the breakup happens by peeling off liquid mass into the gas environment through the surface undulations while the main jet remains relatively steady. Our corresponding jet Weber number is more closely related to the "fiber breakup" regime in that work (8).

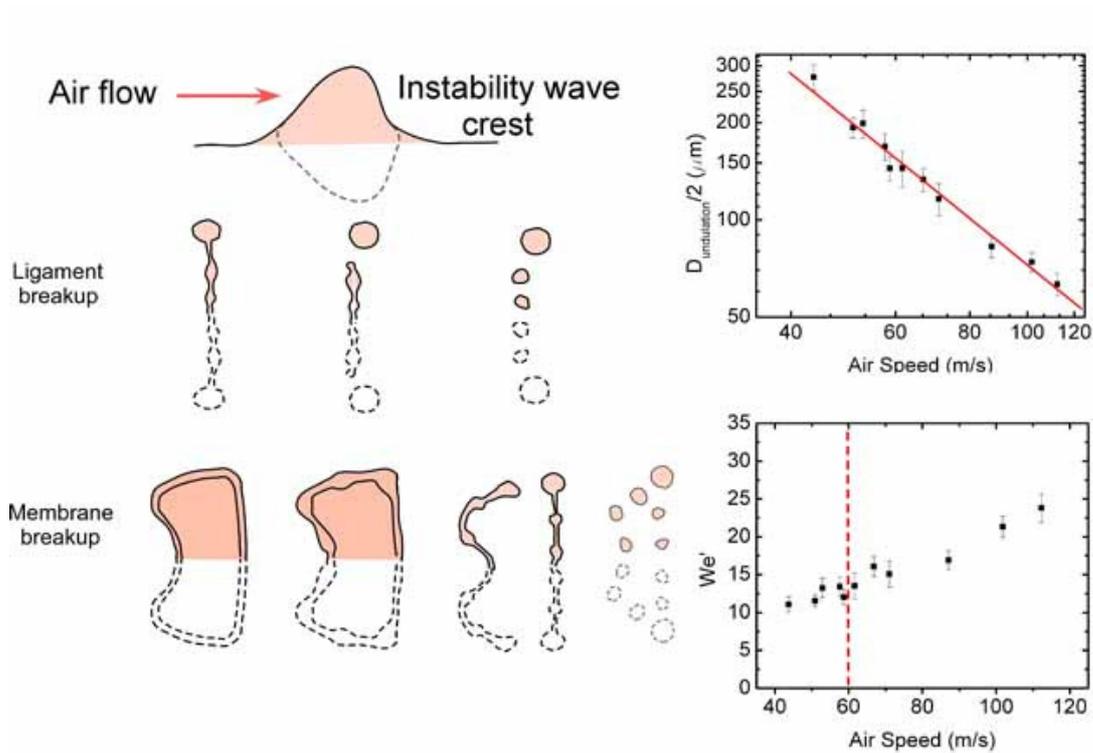

Fig. 4. Left panel shows the schematic of a surface undulation in comparison with a half droplet. The ligament and membrane breakup modes observed in the experiment are illustrated schematically. Right top panel shows the effective $D_{undulation}$ as a function of air speeds. Right bottom panel shows the corresponding $We'$ as a function of air speed. We observe that the transition from ligament- to membrane- mediated breakup happens around $We' \sim 13$ ( $V_g \sim 60.0$ m/s).

The ligament-mediated and membrane-mediated breakups observed in the current study resembles greatly to the droplet breakup in a high-speed air stream. The similarity does not seem to be coincidental. When a droplet breaks in high-speed air stream, there exists a reflection symmetry of the breakup against the meridional plane (4). As shown in Fig.4, the surface crests were exposed to the high speed air flow once formed. Since the surface undulations are normally much smaller in size than the jet itself, the dynamics of them can be treated quite independently from the main jet. As a result, the dynamics of the surface undulation in high speed air flow could be treated as a half liquid sphere. Our experiment reveals that this phenomenological conjecture seems surprisingly self-consistent. To establish this analogy quantitatively, we calculated the effective Weber number ($We' = \rho_g (V_g - V_l)^2 D_{undulation} / \sigma$) of the surface undulations by its effective diameter (estimated by integrating over the breakaway volume $V_{break}$ in a single breakup event), where $D_{undulation}$ is the equivalent half sphere diameter which has same liquid volume with $V_{break}$ ($V_{break} = 2\pi/3 (D_{undulation}/2)^3$).

Since the imaging process is by nature two-dimensional. Careful precautions were taken in estimating $V_{break}$. In the ligament-mediated breakup case, the volume integration is either estimated by adding up all individual droplet volumes assuming that fairly spherical droplets are formed once they relax sufficiently in the air stream, or assuming an axial symmetry around the ligament central axis, and break up the ligament into small regions for integration as illustrated in Fig. 2b and Fig. 2f.

In the membrane-mediated breakup regime the integration is more tricky since the membrane itself can carry substantial mass and most of time there is no obvious symmetry in the system. As a result, the integration is mostly carried out after the membrane bursts when the remaining frame has axial symmetry and the retracting membrane carries negligible mass except at its boundary.

Two results stand out instantly once the effective breakup radii for different air speeds are obtained. Plotting on a log scale, it is clearly that there exist a clean power law between the air speeds and the effective breakup radii $D_{undulation} \sim V_g^{-3/2}$, a similar graph has been obtained by Marmottant *et al.* (2) in which the coefficient was fit to match -1. This observation is very surprising given the highly turbulent and chaotic nature of our breakup process. On the other hand, it implies that the breakup is mostly determined by fundamental parameters like the air speed, liquid surface tension and physical dimensions irrespective of the other details like turbulence.

Additionally, there exists a almost linear relationship between *We'* and $V_g$. The undulations will break up under ligament-mediated breakup mode (*We'* < 13), membrane-mediated breakup mode (*We'* > 13) accordingly, which is in quantitative agreement to the Weber number ranges observed for droplet breakup in high speed air stream (8, 9). This observation also has significant implications since the quantitative match in both the Weber number and breakup phenomenology suggests that two phenomena can be used interchangeably.

We have made the presumption that the surface-undulation-mediated breakup can be treated fairly independent from the main jet dynamics, this presumption is also confirmed by the quantitative study. As is obvious from Fig. 4, when the air speed is fairly high, each breakup event will remove liquid mass with much smaller radius as compared to the water jet diameter, thus the dynamics of the jet can be safely ignored in the first approximation. Only when the air speeds approaches zero, the effective breakup diameter will diverge so the dynamics of the jet has to be included, in which we encounter the familiar jet Rayleigh breakup and jet membrane breakup (8).

If we include previous results on the analogy between the jet breakup and the droplet breakup, it is tempting to speculate a cascade breakup scenario is at work: effective Weber number on different length scales can be defined and the breakup is solely dependent on the effective Weber number. The dynamic Weber number remains the universal criteria in determining the breakup modes.

It is well known that one of the dominant difficulties of the jet breakup research has been the complexity of the processes involved; direct simulation is practically impossible. To simplify the simulation process, sub-models have to be adopted, it is thus the goal of direct simulation and experiments to establish these sub-models (10, 11). In the current letter, we have made a big progress in establishing the intrinsic analogy between the primary breakup of air-assisted liquid jet with its secondary breakup. Ligament-mediated breakup and membrane-mediated breakup are found to be the most important modes of the breakup mechanisms. We speculate that a cascade application of these two breakups is crucial to the understanding of the breakup phenomena.

The authors acknowledge Paul Micheli from Illinois Tool Works Inc. (ITW) and Jin Wang from APS for providing the commercial spray gun system used in the current study and helpful discussions. This work and the use of the APS are supported by the U.S. Department of Energy, Office of Science, Office of Basic Energy Sciences, under Argonne National Laboratory Director's Competitive Grant (LDRD) 2006-023-N0.